\documentclass[conference]{IEEEtran}
\IEEEoverridecommandlockouts
\usepackage{cite}
\usepackage{amsmath,amssymb,amsfonts}
\usepackage{algorithmic}
\usepackage{graphicx}
\usepackage{textcomp}

\newtheorem{theorem}{Theorem}
\newtheorem{corollary}{Corollary}[theorem]
\newtheorem{definition}{Definition}

\def\BibTeX{{\rm B\kern-.05em{\sc i\kern-.025em b}\kern-.08em
    T\kern-.1667em\lower.7ex\hbox{E}\kern-.125emX}}

\graphicspath{{./mats_images/}}

\begin{document}

\title{Compressed Fourier-Domain Convolutional Beamforming for Wireless Ultrasound imaging }
\author{Alon Mamistvalov and Yonina C. Eldar
	\thanks{The work was supported  in part by the Igel Manya Center for Biomedical	Engineering and Signal Processing, as well as the Benoziyo Endowment Fund for the Advancement of Science, the Estate of Olga Klein – Astrachan, and the European Union’s Horizon 2020 research and innovation program under grant No. 646804-ERC-COG-BNYQ; }
	\thanks{A. Mamistvalov and Y. C. Eldar are with the Faculty of Math and CS,	Weizmann Institute of Science, Rehovot, Israel, email: \{alon.mamistvalov, yonina.eldar\} @weizmann.ac.il.}}
\maketitle	
	\begin{abstract}
	Wireless ultrasound (US) systems that produce high-quality images can improve current clinical diagnosis capabilities by making the imaging process much more efficient, affordable, and accessible to users. The most common technique for generating B-mode US images is delay and sum (DAS) beamforming, where an appropriate delay is introduced to signals sampled and processed at each transducer element. However, sampling rates that are much higher than the Nyquist rate of the signal are required for high resolution DAS beamforming, leading to large amounts of data, making transmission of channel data over WIFI impractical. Moreover, the production of US images that exhibit high resolution and good image contrast requires a large set of transducers which further increases the data size. Previous works suggest methods for reduction in sampling rate and in array size. In this work, we introduce compressed Fourier domain convolutional beamforming, combining Fourier domain beamforming, sparse convolutional beamforming, and compressed sensing methods. This allows reducing both the number of array elements and the sampling rate in each element, while achieving high resolution images. Using \emph{in vivo} data we demonstrate that the proposed method can generate B-mode images using 142 times less data than DAS. Our results pave the way towards wireless US and demonstrate that high resolution US images can be produced using sub-Nyquist sampling and a small number of receiving channels.
\end{abstract}


\section{Introduction}
\label{sec:introduction}
\IEEEPARstart{U}{ltrasound} (US) imaging is one of the most common medical imaging methods. It offers a wide range of noninvasive applications, including cardiac, fetal, and breast imaging. In traditional US, imaging is performed by transmitting acoustic pulses along a narrow beam from an array of transducer elements. While propagating, echoes are scattered by acoustic impedance perturbations in the tissue and detected by the same array of transducers. The data collected by the receiving elements is then stored and processed to create an image line in a way called beamforming \cite{van1988beamforming}. In the process of beamforming, the signals are aligned by introducing appropriate time delays and subsequently averaged. Beamforming allows to focus and steer the beam to a desired direction corresponding to the transmission path or  a point in space. This results in signal-to-noise ratio (SNR) improvement together with improved angular localization, which makes it one of the main components in the imaging cycle. 

The standard approach for beamforming is delay and sum (DAS) beamforming \cite{thomenius1996evolution}, \cite{steinberg1992digital} due to its low computational cost and real time capabilities. In DAS beamforming delays are implemented digitally after sampling the signals in order to align the data, before averaging over the channels. To allow high resolution time delays and avoid artifacts caused by the digital implementation of beamforming in time, US signals are typically sampled at a rate 4 to 10 times higher than their Nyquist rate \cite{eldar2015sampling}. 

To generate US images with high resolution and image contrast, the beam pattern of the beamformer should have narrow main lobe and low sidelobes \cite{thomenius1996evolution}, \cite{ranganathan2003novel}. Although being simple for real time applications, DAS suffers from low image resolution and contrast. Increasing the number of array elements can improve image quality when keeping the array pitch below half a wavelength to avoid grating side-lobes \cite{lockwood1998real}. However, a large number of transducer elements, each sampling the signal at high rates, results in an enormous amount of data that needs to be stored and processed, as well as many sampling channels. The large amount of data leads to impractical demands on the hardware and system power when considering portable devices and wireless probes. Therefore, rate reduction, together with image quality enhancement, is of great importance for portable US devices.

Several methods have been studied recently for data, power, and sampling rate reduction. Reduction in sampling rate has been investigated in several studies, based on the combination of compressed sensing (CS) ~\cite{tur2011innovation,wagner2012compressed,chernyakova2014fourier,eldar2012compressed} and sub-Nyquist sampling \cite{eldar2015sampling} by exploiting the finite rate of innovation (FRI) \cite{vetterli2002sampling} structure of the received US signal. It was shown in \cite{wagner2012compressed}, \cite{chernyakova2014fourier} that DAS beamforming can be implemented equivalently in the Fourier domain, leading to frequency domain beamforming, with sampling rates lower than those known in today's commercial US systems. This was later extended to plane-wave imaging \cite{chernyakova2018fourier}. The suggested sub-Nyquist system was implemented in the context of radar \cite{baransky2014sub} based on the ideas of Xampling presented in ~\cite{tur2011innovation,michaeli2011xampling,mishali2011xampling,mishali2011xampling2,gedalyahu2011multichannel}. A CS-based synthetic transmit aperture technique is presented in \cite{liu2016compressed}. This method increases the frame rate by transmitting a small number of randomly apodized plane waves, and uses CS to recover the full channel data. However, none of the methods above considered the reduction in receiving elements.

Several approaches for generating US images using fewer receiving elements were studied in the literature. One example is relying on the analog implementation of sub-aperture and microbeamformers \cite{larson19932}, where part of the beamformation is moved to the probe handle. However, this requires producing expensive integrated circuits with high power consumption and affects image quality \cite{savord2003fully}. Other methods considered the usage of sparse arrays \cite{roux2017validation,yen2000sparse,austeng2002sparse,brunke1997broad} where some of the elements are removed. In \cite{cohen2018sparse} Cohen \emph{et al.} introduced a new beamforming method called convolutional beamforming algorithm (COBA). They showed that by applying COBA, the resulting beam-pattern is equivalent to that of the virtual array given by the sum co-array \cite{liu2017maximally}, \cite{cohen2018optimized}. Therefore, one can use thinned sparse arrays, whose sum co-array is a full uniform linear array (ULA), and result in the same beampattern that would have been obtained using the original full array. Using COBA together with sparse economic array geometries can result in a huge reduction of receiving elements, as shown in \cite{cohen2018sparse} and ~\cite{cohen2020sparse,cohen2019sparse,puente1996fractal,werner1999fractal,werner2003overview,feder2013fractals,falconer2004fractal}. Although achieving reduction in the number of receiving elements, these techniques did not address lowering the sampling rate.

In \cite{mamistvalov2020sparse}, we presented the idea of combining both sampling rate reduction and spatial reduction. It was shown that US signals acquired by a sparse array can be sub-sampled to their effective Nyquist rate, delayed in the frequency domain, and then convolutionally beamformed, resulting in a US image with high image quality but produced from a small data set. However, in \cite{mamistvalov2020sparse} the structure of the convolutionally beamformed signal was not exploited to reduce the sampling rates to sub-Nyquist rates.

The main goal of this paper is to present a beamforming and recovery method that reduces the number of receiving elements while sampling each of the channels at a rate lower than the Nyquist rate. We aim for a beamformer that preserves or improves image quality in terms of resolution and contrast, when compared to DAS beamforming. To achieve this goal, we introduce a compressed frequency domain convolutional beamforing algorithm (CFCOBA) which reconstructs the convolutionally beamformed signal from only a portion of the signal's Fourier coefficients.

We begin by introducing the relationship between the Fourier coefficients of the received signal to those of the convolutionally beamformed signal, based on \cite{chernyakova2014fourier}, \cite{cohen2018sparse}. This relation uses the frequency equivalent of delaying signals in time and sparse arrays with desired sum co-array properties. We then show that the Fourier coefficients of the convolutionally beamformed signal can be calculated efficiently using the fast Fourier transform (FFT), making its implementation in real time possible. The proposed method for calculating the convolutionally beamformed signal Fourier coefficients requires only a portion of the signals bandwidth and uses sparse arrays, resulting in massive data size reduction. To reconstruct the convolutionally beamformed signal, we prove that the signal obeys an FRI model based on the square of the known transmitted acoustic pulse, which enables recovery using known CS methods.

Next, we evaluate the suggested technique on simulated and \emph{in vivo} data of several body parts scanned by different US machines. Using this data we show that US images can be produced without impacting image quality and even improving it when compared to DAS using up to two orders of magnitude less data. The data used for the proposed method is sampled at a rate lower than its effective Nyquist rate, which is typically much lower than its highest frequency. Thus, we illustrate that CFCOBA allows preservation of image quality with up to 142-fold reduction in data size due to the lower sampling rate and efficient sparse arrays used upon reception. Our approach offers significant data size reduction compared to common methods used today, combined with an efficient implementation, which can impact the system size, power consumption, cost, and mobility, making wireless US imaging feasible.

The rest of the paper is organized as follows. Section \ref{sec: USImageRec} discusses several beamforming methods such as DAS, Fourier domain beamforming, and COBA, together with examples of sparse arrays. Section \ref{sec: FDBF_not_compressed} presents our approach for combining data reduction in the acquiring array size and in the sampling rate. We consider further reduction in sampling rate in Section \ref{sec: CFCOBA} and reconstruction of the signal using CS methods. The performance of the suggested technique is evaluated in Section \ref{sec: EvResults} using phantom and \emph{in vivo} scans. The paper is concluded in Section \ref{sec: conclusion}.

Table \ref{table: notation} summarizes notation used throughout the paper.

\begin{table*}[h!]
	\centering
	\caption{List of Notation}
	\label{table: notation}
	\begin{tabular}{||c|c||}
		\hline
		$M$ & Transducer array geometry\\		
		$|M|$ & Size of array $M$\\
		$\min(M), \max(M)$ & Minimum and maximum element of array $M$\\
		$\boldsymbol{a}$ & Vector\\
		$\boldsymbol{A}$ & Matrix\\
		$(\cdot)^*$ & Complex conjugate\\
		$*, \underset{s}{*}$ & discrete linear convolution over Fourier coefficients and discrete linear convolution over spatial dimension\\
		$\phi_m(t)$ & Signal received at $m$th transducer element\\
		$\hat{\phi}_m(t;\theta)$ & Dynamically delayed received signal used for beamforming in direction $\theta$\\
		$\Phi(t;\theta)$ & Beamformed signal at direction $\theta$\\
		$u_m(p;\theta)$ & Normalized delayed signal at channel $m$\\
		$T$ & Time duration of the receiving signal\\
		$c_m[k]$ & $k$th Fourier coefficient of the received signal at channel $m$\\
		$c[k]$ & $k$th Fourier coefficient of the beamformed signal\\
		$\hat{c}_m[k]$ & $k$th Fourier coefficient of the delayed signal at channel $m$\\
		$f_s$ & Sampling rate for traditional beamforming\\
		$N_{st} = \lfloor T \cdot f_s$ $\rfloor$ & Number of samples required for traditional beamforming\\
		$\beta$  & Effective band-pass bandwidth of the recieved signal\\
		$B$ & Cardinality of $\beta$\\
		$f_{sEN} = B/T$ & Nyquist sampling rate of effective band-pass bandwidth of the received signal\\
		$f_{sN}$ & Sub Nyquist sampling rate\\
		$\beta_{sN}$ & Subset of $\beta$ used for sub-Nyquist sampling\\
		$B_{sN}$ & Cardinality of $\beta_{sN}$\\
		$N_{sN} = 2B_{sN}-1$ & Number of time samples corresponding to sub Nyquist rate\\
		\hline			
	\end{tabular}
\end{table*}

\section{Ultrasound Beamforming Techniques}
\label{sec: USImageRec}
In most US imaging systems, the US image is built line by line for each direction $\theta$, using multiple transducer elements to transmit and receive acoustic pulses. In that way, beamforming can be performed both during transmission and reception. In transmission, a pulse is generated and transmitted by an array of transducer elements. The pulse transmitted by each element in the array is delayed and scaled so that their sum  creates a directional beam propagating at a certain direction through the tissue. A whole sector is radiated by subsequently transmitting at different angles. The transmission parameters per angle are calculated offline and saved in tables, which makes real-time computational complexity in the transmit negligible. Beamforming in reception, however, is much more challenging. It involves dynamically delaying signals received at each of the transducer elements and averaging them. By averaging the delayed signals, the SNR of the beamformed signal is enhanced and a line in the image is formed. In the next subsections, we review several receive beamforming methods that allow reducing the number of samples or the number of receiving antenna elements.
\subsection{Time Domain Beamforming}
\label{sec: TDBF}
Consider a uniform linear array, $M$, comprised of $|M| = 2N-1$ transducer elements aligned along the x-axis. The imaging cycle starts at $t=0$, when the pulse is transmitted by each transducer element, resulting in a beam propagating at direction $\theta$ through the tissue. The energy is scattered by reflectors and the echoes are received by all elements at times that depend on their location. Denote by $\phi_m(t)$ the signal received by the $m$th element. 
Beamforming involves averaging the signals received by different array elements while compensating using time shifts for alignment due to the differences in arrival time. 
The delay that needs to be applied is defined by the speed of sound in the tissue, $c$, and by the distance between the reference element, $m_0$, and the receiver, $\delta_m$. Applying an appropriate delay to $\phi_m(t)$, the reflection at each receiving element is aligned to the reference point, resulting in $\hat{\phi}_m(t;\theta)$ which is the aligned signal \cite{wagner2012compressed},\cite{jensen1999linear}:
\begin{align} \label{delayeDsignal_DAS}
&\hat{\phi}_m(t;\theta) = \phi_m(\tau_m(t,\theta)), \nonumber \\
&\tau_m = \dfrac{1}{2}(t+\sqrt{t^2-4(\delta_m/c)t\sin \theta +4(\delta_m/c)^2}).
\end{align}
These signals are then averaged to form the beamformed signal for direction $\theta$:
\begin{equation} \label{DAS_eq}
\Phi_{DAS}(t;\theta)= \dfrac{1}{|M|} \sum^{|M|}_{m=1} \hat{\phi}_m(t;\theta).
\end{equation}

In practice, DAS beamforming is performed digitally. The applied delays are on the order of nanoseconds which results in a sampling rate that can be as high as hundreds of megahertz \cite{demuth1977frequency}, a requirement that is impractical. Therefore, US data signals are sampled at lower rates, on the order of tens of megahertz and fine delay resolution is obtained by subsequent digital interpolation which adds an additional computational load. Yet, these lower rates are much higher than the Nyquist rate of the signal which is twice its' bandwidth \cite{eldar2015sampling}. A well known rule of thumb is that the sampling rate of the signal should be 4 to 10 times the transducer central frequency. These high sampling rates, together with the large number of array elements used, results in a huge amount of data which makes US channel data transmission over WIFI impossible.

\subsection{Frequency Domain Beamforming}
\label{sec: FDBF}
Frequency domain beamforming was suggested in \cite{chernyakova2014fourier} in order to reduce the sampling rate. Chernyakova \emph{et al.} showed the equivalence of performing beamforming in time and in the Fourier domain. It was then shown that beamforming can be performed efficiently using a small number of Fourier coefficients of the received signals. Let $c[k]$ denote the $k$th Fourier series coefficient of the beamformed signal and $\hat{c}_m[k]$ the $k$th Fourier series coefficient of the delayed signal at channel $m$. Based on \eqref{DAS_eq} and the linearity of the Fourier transform, it can be seen that
\begin{equation}\label{FDBF_eq_delay_BF}
c[k] = \dfrac{1}{|M|} \sum^{|M|}_{m=1} \hat{c}_m[k],
\end{equation}
where $\hat{c}_m[k]$ is defined as
\begin{equation}
\hat{c}_m[k] = \dfrac{1}{T}\int_{0}^{T} I_{[0,T_B(\theta))}(t)\hat{\phi}_m(t;\theta)e^{\frac{-2\pi j}{T}kt }dt.
\end{equation}
Here $I_{[a,b)}$ is the indicator function equal to $1$ when $a\leq t < b$. The beam is supported on $\left[ 0, T_B(\theta)\right)$, where $T_B(\theta) < T$ is defined in \cite{wagner2012compressed} and $T$ is the pulse penetration depth.

According to \cite{wagner2012compressed} and \cite{chernyakova2014fourier}, the Fourier coefficients of the delayed signal can be expressed as
\begin{equation} \label{eq_dist_fun_Fourier}
\hat{c}_m[k] = \sum^{\infty}_{n = -\infty} c_m[k-n]Q_{k,m;\theta}[n],
\end{equation}
where $c_m[k]$ are the Fourier coefficients of the received signals in each channel before delay. The variables $Q_{k,m;\theta}[n]$ are the Fourier coefficients of a distortion function which is determined solely by the geometry of the imaging setup and can be computed offline once and stored in memory using a look up table (LUT) for real time calculations. By using the Fourier coefficients of the distortion function, we effectively transfer the beamforming delays to the frequency domain. The summation in \eqref{eq_dist_fun_Fourier} can be replaced by a relatively small finite summation due to the decay properties of $\left\lbrace Q_{k,m;\theta}[n]\right\rbrace $ and the fact that most of the energy of this set is centered around the DC component. Thus, $\left\lbrace Q_{k,m;\theta}[n]\right\rbrace$ decays rapidly for $n < -N_1, \; n > N_2$ where $N_1, N_2 \in \mathbb{N}$, leading to
\begin{equation} \label{eq_dist_fun_Fourier_notFinal_finite}
\hat{c}_m[k] = \sum_{n=-N_1}^{N_2} c_m[k-n]Q_{k,m;\theta}[n].
\end{equation}
Combining \eqref{eq_dist_fun_Fourier_notFinal_finite} and \eqref{FDBF_eq_delay_BF} yields frequency domain beamforming (FDBF):
\begin{equation} \label{eq_dist_fun_Fourier_finite}
c[k] = \dfrac{1}{|M|} \sum^{|M|}_{m=1} \sum_{n=-N_1}^{N_2} c_m[k-n]Q_{k,m;\theta}[n].
\end{equation}
Appropriate zero padding and applying an inverse Fourier transform to $\left\lbrace c[k]\right\rbrace $, results in the time domain beamformed signal.

The importance of FDBF is that it only requires the nonzero Fourier coefficients of the received signal. Those coefficients are obtained from sub Nyquist samples of the signal at each receiving element. Following \cite{chernyakova2014fourier} we know that the beamformed signal will contain at most $(B+N_1+N_2)$ nonzeros frequency components, where $B$ is the cardinality of the set of $k$s for which $c_m[k]$ is nonzero. In practice, due to the fast decaying property of the distortion function, $B \gg N_1, N_2$ implies that the bandwidth of the beamformed signal equals $B$. Hence, to perform beamforming in frequency, we need only a portion of the Fourier coefficients of the signal. To obtain them, we use the Xampling mechanism proposed in \cite{tur2011innovation}. The implementation of this mechanism is discussed in \cite{baransky2014sub}. The output is sampled at its effective Nyquist rate and the required Fourier coefficients are the Fourier transform of the output. This yields a data size reduction of $N_{st}/B$.

\subsection{Convolutional Beamforming}
\label{sec: COBA}
Substantial data reduction can also be achieved by the COnvolutional Beamforming Algorithm (COBA), \cite{cohen2018sparse}, \cite{cohen2020sparse}, which produces images, at least as good as those obtained by DAS, using fewer array elements. The key idea is that the effective beampattern obtained by COBA is equivalent to the beampattern that would have been obtained using the sum co-array of the given physical array \cite{hoctor1990unifying}, which produces images with better resolution and contrast. 

Let $\phi_m(p,\theta)$ denote the delayed signal at channel $m$, sampled at sampling intervals $T_s = \frac{1}{f_s}$, where $p$ stands for the $p$th sample. The beamformed signal in COBA is
\begin{equation} \label{COBA_def}
\Phi_{COBA}(p;\theta) = \sum^{N-1}_{n = -(N-1)} \sum^{N-1}_{m = -(N-1)} u_n(p;\theta)u_m(p;\theta),
\end{equation}
where
\begin{equation} \label{signal_norm}
u_m(p;\theta) = e^{j\angle\hat{\phi}_m(p;\theta)} \sqrt{|\hat{\phi}_m(p;\theta)|},
\end{equation}
with $\angle\hat{\phi}_m(p,\theta)$ and $|\hat{\phi}_m(p,\theta)|$ being the phase and the magnitude of $\hat{\phi}_m(p,\theta)$, respectively. The operation in \eqref{signal_norm}, ensures that the amplitude of each product in \eqref{COBA_def} will be on the same order of the received signal at each channel. The relationship in \eqref{COBA_def} can be written equivalently as
\begin{equation} \label{COBA_efficient1}
\Phi_{COBA}(p;\theta) = \sum^{2(N-1)}_{n = -2(N-1)}s_n(p;\theta),
\end{equation}
where
\begin{equation} \label{notation_conv_u_s}
s_n(p;\theta) = \sum_{i,j\in U,i+j=n} u_i(p;\theta)u_j(p;\theta),
\end{equation}
with $ n = -2(N-1),...,2(N-1)$. Therefore, the vector $\boldmath{s}$, whose entries are $s_n$ can be calculated by
\begin{equation}\label{conv_signal}
s(p;\theta) = u(p;\theta)\underset{s}{*}u(p;\theta).
\end{equation}
Convolution is obtained by zero padding $u$ to length $2|M|-1$. Therefore, we can equivalently apply lateral convolution on the delayed signals, and then sum the elements.

To achieve a beampattern with desired properties we examine the beampattern of the DAS beamformer \cite{cohen2018sparse}
\begin{equation} \label{DAS_BP}
H_{DAS}(\theta) = \sum_{n=-(N-1)}^{N-1} \exp\left( {-j\omega_0\frac{\delta_n \sin\theta}{c}}\right) ,
\end{equation}
where $\omega_0$ is the central frequency of the transducer and $\delta_n = n\delta$, with $\delta$ being the distance between two consecutive array elements. Based on the definition of COBA in \eqref{COBA_def}, we see that $H_{COBA} = H_{DAS}H_{DAS}$, which leads to
\begin{equation}
H_{COBA}(\theta) =  \sum_{n,m=-(N-1)}^{N-1} \exp\left( {-j\omega_0\frac{\delta \sin\theta}{c}(n+m)}\right) .
\end{equation}
Following \cite{cohen2018sparse}, this summation can be written as a single polynomial
\begin{equation} \label{COBA_final_BP}
H_{COBA}(\theta) =  \sum_{n=-2(N-1)}^{2(N-1)} a_n\exp\left( {-j\omega_0\frac{\delta \sin\theta}{c}n}\right),
\end{equation}
where ${a_n}$ are intrinsic apodization weights given by $\boldsymbol{a} = \mathbb{I}_M*\mathbb{I}_M$. Here $\mathbb{I}_M$ is a binary vector whose $m$th entry is 1 if $m \in M$. 

\begin{definition}
	\label{def: sumset}
	Sum co-array: Consider a linear array $M$. We define the set
	\begin{equation}
	\tilde{S}_M = \left\lbrace  n+m: n,m\in M \right\rbrace.
	\end{equation}
	The sumset of the set $M$ is $S_M$ and is defined to consist of only the distinct elements of $\tilde{S}_M$. The array with elements located at $n\delta$ and $n\in S_I$, is the sum co-array of $M$. 
\end{definition}
Equation \eqref{COBA_final_BP} can be thought of as the beam pattern of the DAS beamformer operating on the sum co-array. The sum co-array is larger than the original array which leads to improved imaging performance thanks to the effective beampattern. Therefore, a thinned array, comprised of the original array after removing some of its elements, can be used to actually reduce the data size. By preserving the desired sum co-array, the beampattern is not changed and might be improved.

To obtain thin arrays with desirable properties, \cite{cohen2020sparse} suggested the use of fractal arrays ~\cite{puente1996fractal,werner1999fractal,werner2003overview,feder2013fractals,falconer2004fractal}. Fractal arrays are defined recursively by
\begin{align} \label{fractal}
&W_0 = {0}, \nonumber \\
&W_{r+1} = \cup_{n\in \mathbb{G}}(W_r + nL^r), \; r\in \mathbb{N},
\end{align}
where the array $\mathbb{G}$ is the \emph{generator} array in fractal terminology, with $\min(\mathbb{G}) = 0$. The translation factor $L$ is given by $L = 2\max(\mathbb{G})+1$ where $r$ is the array order. 
The resulting array, $W$, is composed of spatially arranged copies of $\mathbb{G}$. This choice leads to very thin arrays with desirable properties, and co-arrays that include the ULA of size $|M|$. 

Using thinned arrays leads to reduction in power and data rates since less data is sampled, processed and stored.

\section{Frequency Domain COBA} \label{sec: FDBF_not_compressed}
In this section, we follow \cite{mamistvalov2020sparse} and show that FDBF can be combined with COBA. This leads to an efficient beamforming method using a small number of array elements, each sampled at the effective Nyquist rate of the signal, without impacting the image quality compared to DAS.
\subsection{Implementation}
Consider a ULA of desired aperture $|M|$ and a given array geometry $U \subseteq M$ corresponding to a desired array following Section \ref{sec: COBA}. The signal $\phi_m(t;\theta)$ is acquired by the receiving element $m$. Let $c_m[k]$ be the $k$th Fourier series coefficient of the received signal at channel $m$
\begin{equation}
c_m[k] = \dfrac{1}{T}\int_{0}^{T} I_{[0,T_B(\theta))}(t)\phi_m(t;\theta)e^{\frac{-2\pi j}{T}kt }dt.
\end{equation}
As stated in \cite{chernyakova2014fourier}, a typical ultrasound signal has one main band of energy, of bandwidth $B$. The energy outside this band is much lower, hence the sampling rate of the signal is set to achieve a consecutive set of the Fourier coefficients of the received signals, $\beta$, such that $|\beta|=B$. This implies that the sampling rate of the signal is dictated by the bandwidth of the signal, and denoted by $f_{sEN} = B/T$, with $T$ defined in \ref{sec: FDBF}.

Next, as shown in \eqref{eq_dist_fun_Fourier_notFinal_finite}, by multiplying the Fourier coefficients of the distortion function and the Fourier coefficients of the received signals elementwise, an appropriate delay is applied to the received signals, obtaining the Fourier coefficients of the delayed signal.
The frequency domain delayed (FDD) signal, $\hat{\phi}^{FDD}_m(p;\theta)$, is obtained by applying an inverse Fourier transform on \eqref{eq_dist_fun_Fourier_notFinal_finite}. Here $p$ denotes the $p$th sample of the signal on the time grid of the delayed signal, $p = 1...N_{st}$. By plugging $\hat{\phi}^{FDD}_m(p;\theta)$ into \eqref{signal_norm} we get
\begin{equation}
u^{FDD}_n(p;\theta) = u^{TDD}_n(p;\theta).
\end{equation}
We refer to time domain delayed as TDD, which holds for a signal delayed using traditional delaying as shown in \eqref{delayeDsignal_DAS}. The key idea is that the delay can be achieved regardless of the array geometry. Having obtained the equivalence of the delayed signals in time and frequency, we can easily proceed in applying COBA.

Finally, we plug both the time domain and the frequency domain delayed signals into \eqref{COBA_def} and proceed according to Section \ref{sec: COBA}, which results in equivalence of the convolutional beamforming applied to TDD signal and FDD signal.

\subsection{Data reduction}
Data size reduction and power reduction are two of the main challenges encountered when trying to implement a wireless US imaging system. Power and data size reduction are achieved automatically by reducing the number of receive channels. Moreover, the modulated transmitted pulse has one main band of energy, thus, only a consecutive set of Fourier coefficients that is contained in this band needs to be obtained, using the Xampling mechanism. Therefore, by combining COBA and FDBF, one can sample using a sparse set of array elements each sampled at the Nyquist rate of the signal. The resulting data size reduction is therefore $\frac{|M|}{|\tilde{U}|} \frac{N_{st}}{B}$, where $|M|, |\tilde{U}|$ are the original and the thinned arrays size, respectively, $N_{st}$ is the number of samples traditionally needed for DAS and $B$ is the actual signal bandwidth.

\section{Compressed Frequency Domain COBA}
\label{sec: CFCOBA}
We now consider reconstruction of the convolutional beamformed signal from partial frequency data. To this end, we first derive a relation between the Fourier coefficients of the convolutionally beamformed signal and the Fourier coefficients of the received signals. Then, we examine the FRI structure of the convolutionally beamformed signal and prove that based on the FRI model we can reconstruct the convolutionally beamformed signal from only a portion of its Fourier coefficients.

\subsection{Exploiting the Fourier Coefficients Relationship}
Here we derive the relation between the Fourier coefficients of the received signal at each channel of $U$, the desired array geometry defined in Section \ref{sec: COBA}, and those of the convolutionally beamformed signal. The calculation is done for selected $\theta$ and for signals that correspond to FDD; those subscripts are not written explicitly for brevity.
We aim to reduce the sampling rate further than proposed in \cite{mamistvalov2020sparse}, by obtaining a set of consecutive Fourier coefficients, $\beta_{sN} \subseteq \beta$. The set $\beta_{sN}$ is acquired using an appropriate filter \cite{tur2011innovation}
\begin{equation}
\phi_m(n_s) = \int_{-\infty}^{\infty} \phi_m(t)f_r(t-n_sT_{sN})dt,
\end{equation}
where
\begin{equation}
f_r(t)  = \sum_{l=-r}^{r}f(t+lT),
\end{equation}
$r$ is a constant determined by the support of the transmitted pulse, $T$ is the duration of the received signals, and $T_{sN}$ is the sampling period for sub-Nyquist sampling.
The filter, $f(t)$, satisfies the following frequency response
\begin{align}
F(\omega) = 
\begin{cases}
0, \: \text{if} \:\: \omega = \dfrac{2\pi k}{T},\;  k\notin \beta_{sN} \\
1, \: \text{if} \:\: \omega = \dfrac{2\pi k}{T}, \; k\in \beta_{sN} \\
arbitrary, \:\: \text{else}.
\end{cases}
\end{align} 	 	

Let $N_{sN}$ be the number of samples acquired by sub-Nyquist sampling of the received signal. We define the compressed frequency domain signal by
\begin{equation} \label{COBA_no_noramlization}
\hat{\Phi}(n_s)_{CFCOBA} = \sum_{n \in U} \sum_{m \in U} \hat{\phi}_n(n_s)\hat{\phi}_m(n_s),
\end{equation}
where $n_s=0,...,N_{sN}-1$ are the discrete time samples, and $\hat{\phi}_m(n_s)$, $\hat{\phi}_n(n_s)$ are the delayed signals at channels $m$ and $n$ respectively, with the delay applied in the frequency domain. Each of the delayed signals has $B_{sN} = |\beta_{sN}|$ Fourier coefficients that are not zero, dictated by the filters width. The Fourier coefficients of the acquired signals are zero padded to length $N_{sN} = 2B_{sN}-1$. Based on the multiplication in the time domain, the signal $\hat{\Phi}(n_s)_{CFCOBA}$ has $2B_{sN}-1$ non-zero Fourier coefficients.

Next, we write the Fourier coefficients of the convolutionally beamformed signal based on the $N_{sN}$ samples acquired
\begin{equation}
\hat{c}[k]_{CFCOBA} = \sum_{l \in U}\sum_{m \in U} \sum_{n_s=0}^{N_{sN}-1} \hat{\phi}_l(n_s) \hat{\phi}_m(n_s)\exp\left( \dfrac{-2\pi i}{N_{sN}}kn_s\right).
\end{equation}
By plugging the Fourier series coefficients of the delayed signal at each channel, we get
\begin{align}
\hat{c}[k]_{CFCOBA}& = \sum_{l \in U}\sum_{m \in U} \sum_{n_s=0}^{N_{sN}-1} \sum_{p = 0}^{N_{sN}-1}\hat{c}_m[p]\exp\left( \dfrac{2\pi i}{N_{sN}}pn_s\right) \\  \nonumber &\sum_{q = 0}^{N_{sN}-1}\hat{c}_l[q]\exp\left( \dfrac{2\pi i}{N_{sN}}qn_s\right)  \exp\left( \dfrac{-2\pi i}{N_{sN}}kn_s\right), 
\end{align}
where the number of nonzero elements of $\{c_m[p]\}, \{c_l[q]\}$ is $B_{sN}$. Using the fact that 
\begin{align}
\sum_{n_s=0}^{N_{sN}-1} \exp\left( \dfrac{2\pi i}{N_{sN}}(p+q-k)n_s\right)= 
&\begin{cases}
N_{sN},      & \text{if} \quad  q+p = k\\
0,      & \text{else},
\end{cases}
\end{align}
we get
\begin{align} \label{conv_almost_final}
\hat{c}[k]_{CFCOBA} = &N_{sN}\sum_{l \in U}\sum_{m \in U} \sum_{p+q = k} \hat{c}_m[p]\hat{c}_l[q]   \nonumber \\
=&N_s\sum_{l \in U}\sum_{m \in U} (\hat{c}_m*\hat{c}_l)[k],
\end{align}
where the last equation is obtained by setting $q = k-p$. Hence, the resulting vector $\hat{c}[k]_{CFCOBA}$ has support size $2B_{sN}-1 = N_{sN}$, and $k \in S_{\beta_{sN}}$ where $S_{\beta_{sN}}$ is the sumset of $\beta_{sN}$, following Definition \ref{def: sumset}.

Finally, plugging \eqref{eq_dist_fun_Fourier_notFinal_finite} into \eqref{conv_almost_final}, we get the Fourier coefficients of the convolutionaly beamformed signal
\begin{align} \label{final_conv}
&\hat{c}[k]_{CFCOBA} = N_{sN}\sum_{l \in U}\sum_{m \in U} \\ \nonumber
& \left(  \sum_{w=-N_1}^{N_2} c_m\left[ k-w \right] Q_{k,m}\left[ w\right] 
*\sum_{h=-N_1}^{N_2} c_l\left[ k-h \right] Q_{k,l}\left[ h\right]\right),
\end{align}
where the length of the sequence $c_m\left[ k-w \right]$ is $N_{sN}$ with $B_{sN}$ elements that are not zero. 
To efficiently calculate \eqref{final_conv}, we follow \cite{cohen2018sparse} noticing that \eqref{COBA_no_noramlization} can be written as
\begin{align} \label{COBA_efficient}
\Phi_{CFCOBA}(n_s) = &\sum_{l \in S_{U}} \sum_{m, v \in U:m+v=l } u_m(n_s)u_v(n_s) \nonumber \\
= &\sum_{l \in S_{U}} (u(n_s) \underset{s}{*} u(n_s))_l,
\end{align}
where $S_{U}$ is the sum co-array of $U$. Due to the linear operations in the temporal dimension that led to \eqref{final_conv}, we can plug the results into \eqref{COBA_efficient} leading to
\begin{equation} \label{final_CFCOBA}
\hat{c}[k]_{CFCOBA} = N_{sN} \sum_{l \in S_{U}} (\hat{c} \underset{s}{*}* \hat{c})_l[k],
\end{equation}
which denotes two dimensional convolution operation, one over the temporal dimension and one over the spatial dimension. This calculation can be easily calculated using IFFT and appropriate zero padding based on the convolution theorem, which leads to an efficient implementation with run time complexity of order $O(N_{sN} \log N_{sN})$.

The same derivation can be applied to the convolutionally beamformed signal, sampled at the traditional sampling rate for DAS beamforming. The signal is defined by
\begin{equation} \label{target2}
\hat{\Phi}(n_{st})_{COBA} = \sum_{n \in U} \sum_{m \in U} \hat{\phi}_n(n_{st})\hat{\phi}_m(n_{st}),
\end{equation}
where $n_{st} \in \left\lbrace 0,...,N_{st}-1 \right\rbrace $ is the $n_{st}$th sample of the delayed signal, and the signals are delayed in the frequency domain based on \eqref{eq_dist_fun_Fourier_finite}. In this case, the size of the set of the nonzero Fourier coefficients is $B \gg B_{sN}$ and $N_{st} \gg N_{sN}$.
This results in
\begin{align} \label{final_conv_traditional}
&\hat{c}[k]_{COBA} = N_{st}\sum_{l \in U}\sum_{m \in U} \\ \nonumber
&\left(  \sum_{w=-N_1}^{N_2} c_m\left[ k-w \right] Q_{k,m}\left[ w\right] 
*\sum_{h=-N_1}^{N_2} c_l\left[ k-h \right] Q_{k,l}\left[ h\right]\right),
\end{align}
where $\{c_m[k]\}_{k=0}^{N_{sN}-1} \subset \{c^{st}_m[k]\}_{k=0}^{N_{st}-1}$ and $c_m[k] = c^{st}_m[k], \: \forall k \in \beta_{sN}$. The resulting sub sampled signal Fourier coefficients thus satisfy
\begin{equation}
\hat{c}^{sN}[k]_{CFCOBA} = \hat{c}^{st}[k]_{COBA}, \: \forall k \in S_{\beta_{sN}}.
\end{equation}

We next discuss reconstruction based on this partial set of Fourier coefficients.
\subsection{FRI Structure Derivation} 	
We first show that the convolutionally beamformed signal follows an FRI model. 	 	
\begin{theorem}
	Let $\hat{\Phi}(t)_{COBA}$ be the convolutionally beadformed signal defined in \eqref{target2} over continuous time. For any acquiring array, $U$, the convolutionally beamformed signal can be modeled as an FRI signal, i.e. it has the following structure:
	\begin{equation} \label{target_FRI}
	\hat{\Phi}(t)_{COBA} = \sum_{s = 1}^{S} b_sg(t-t_{s}),
	\end{equation}
	where $S$ is the number of scattering elements in the tissue in certain direction $\theta$, $g(t) = h^2(t)$ where $h(t)$ is the transmitted pulse shape, $\left\lbrace b_s \right\rbrace, \left\lbrace t_s \right\rbrace$ are the unknown amplitudes of the reflections and the times at which the reflection from the $s$th scatterer arrived at the receiving element, respectively. 
\end{theorem}

\begin{IEEEproof}
	From \cite{wagner2012compressed} we know that the delayed signal at each element has the form
	\begin{equation} \label{FRI_channels}
	\hat{\phi}_m(t) = \sum_{s = 1}^{S} a_{s,m}h(t-t_{s}).
	\end{equation}
	Substituting \eqref{FRI_channels} into \eqref{target2} we get
	\begin{equation} \label{COBA_channel_FRI}
	\Phi_{COBA}(t) = \sum_{l \in U} \sum_{m \in U}\Big (\sum_{s = 1}^{S} a_{s,l}h(t-t_{s}) \Big) \Big (\sum_{s' = 1}^{S} a_{s',m}h(t-t_{s'}) \Big).
	\end{equation}
	Next, we assume that $h(t)$ is supported on a compact interval $[0, \Delta), \Delta > 0$, meaning that $h(t-t_{s})$ is supported on $[t_{s}, t_{s} + \Delta)$. We also assume that $t_{s} \gg \Delta$ so that in \eqref{COBA_channel_FRI}, all pairwise multiplications of $h(t-t_{s})$ that involve $s\neq s'$ are zero. Thus, the beamformed signal is of the form
	\begin{align} \label{COBA_FRI}
	\Phi_{COBA}(t) =& \sum_{l \in U} \sum_{m \in U} \Big (\sum_{s = 1}^{S} a_{s,l}a_{s,m}h^2(t-{s})\Big ) \nonumber \\
	=&\sum_{s = 1}^{S} \Big (\sum_{l \in U} \sum_{m \in U} a_{s,l}a_{s,m} \Big ) h^2(t-t_{s})  \nonumber \\ 
	=&\sum_{s = 1}^{S} {b}_s g(t-t_{s}) 
	\end{align}
	where $g(t) = h^2(t)$.
\end{IEEEproof}

Using the FRI structure of the convolutionally beamformed signal we now address its reconstruction.
\begin{corollary} \label{cor:1}
	Let $\left\lbrace  \hat{c}[k]_{CFCOBA} \right\rbrace_{k \in S_{\beta_{sN}}} $ be the set of size $2B_{sN}-1 > 2S$ nonzero consecutive Fourier coefficients of the compressed frequency domain convolutionally beamformed algorithm (CFCOBA) signal defined in \eqref{COBA_no_noramlization}. Based on the FRI model, this signal can be recovered from a partial set of its Fourier coefficients, by solving
	\begin{equation} \label{opt_claim}
	\boldmath{\hat{c}}_{CFCOBA} = \boldsymbol{GVb},
	\end{equation}
	where $\hat{\boldsymbol{c}}_{CFCOBA}$ is a vector of size $2B_{sN}-1$ with the nonzero $\left\lbrace  \hat{c}[k]_{CFCOBA} \right\rbrace _{k \in S_{\beta_{sN}}} $ as its entries, $\boldsymbol{G}$ is a diagonal matrix of size $(2B_{sN}-1) \times (2B_{sN}-1)$ with $G(\frac{2\pi k}{T})$ on its diagonal, $\boldsymbol{V}$ is a Fourier matrix of size $2B_{sN}-1 \times S$ with $ (k,s) $th element $e^{-j\left( (2\pi)/T\right) kt_{s}}$, and $\boldsymbol{b}$ the $S$-length vector with the amplitudes, $\left\lbrace b_s \right\rbrace $, as its entries.
\end{corollary}
\begin{IEEEproof}
	We begin by noticing that the signal is completely defined by $2S$ unknown parameters which are the amplitudes, $\{ {b}_s \}$ and the delays, $\{ t_{s} \}$. For a duration $T$, the Fourier series expansion of \eqref{COBA_FRI} can be written as
	\begin{align} \label{FRI_Fourier}
	\hat{c}[k]_{CFCOBA} &= \frac{1}{T}\int_{0}^{T}\sum_{s = 1}^{S} b_sg(t-t_{s})dt \\ \nonumber
	&= G\left( \frac{2\pi k}{T}\right) \sum_{s = 1}^{S}b_se^{-j\left( (2\pi)/T\right) kt_{s}},
	\end{align}
	where $\hat{c}[k]_{CFCOBA}$ are the Fourier coefficients of the convolutionally beamformed signal, and $G(\omega)$ is the continuous time Fourier transform of $g(t)$. Writing \eqref{FRI_Fourier} in matrix form results in \eqref{opt_claim}. This problem is invertible as long as $2B_{sN}-1 > S$ and the time delays $t_s \neq t_{s'},\: \forall s \neq s'$. The formulation in \eqref{opt_claim} is a standard spectral analysis problem and can be solved for the unknown parameters $\{ t_s,b_s \}_{s=1}^S$, using methods like the annihilating filter \cite{stoica1997introduction}. 
\end{IEEEproof}

In practice, the recovery problem can solved using CS methods as will be discussed next.

\subsection{Reconstruction using Compressed Sensing }
\label{sec: CS_reconstruction}

To address recovery using CS methods we begin with \eqref{FRI_Fourier}.
By quantizing the delays with step $T_s = \frac{1}{f_s}$, such that $t_s = q_sT_s$, and letting $N_{st} = \lfloor T/T_s \rfloor$, the Fourier coefficients can be written as
\begin{equation} \label{FRI_Fourier_quant}
\hat{c}[k]_{CFCOBA} = G\left( \frac{2\pi k}{T}\right) \sum_{s = 0}^{N_{st}-1}\tilde{b}_se^{-j\left( (2\pi)/N_{st}\right) ks}.
\end{equation} 
We define the vector $\boldsymbol{\tilde{b}}$ of length $N_{st}$ to consist of
\begin{align} \label{b_quant}
\tilde{b}_s =
\begin{cases}
{b}_s   ,& \text{if} \quad  s=q_s.\\
0   ,& \text{else.} 
\end{cases}
\end{align}
The recovery problem then reduces to determining the $S$-sparse vector $\boldsymbol{\tilde{b}}$ from
\begin{equation} \label{CS_problem}
\boldmath{\hat{c}_{CFCOBA}} = \boldsymbol{GD\tilde{b}} = \boldsymbol{A\tilde{b}},
\end{equation}
where $\boldsymbol{D}$ is a $(2B_{sN}-1) \times N_{st}$ matrix, formed by taking the set $S_{\beta_{sN}}$ of rows from an $N_{st}\times N_{st}$ FFT matrix.	This formulation is a classic CS problem and can be solved using many CS techniques. Choosing $N_{sN} \geq CL(\log N_{st})^4$ rows uniformly at random for some constant $C > 0$, the matrix $\boldsymbol{A}$ obeys the RIP with high probability \cite{rudelson2008sparse}.

In practice, due to speckle, the coefficient vector $\boldsymbol{b}$, defined in \eqref{b_quant}, is only approximately sparse. To reconstruct $\boldsymbol{b}$ we use the $l_1$ norm, leading to
\begin{equation} \label{optimization}
\min _{\tilde{b}} ||\boldsymbol{\tilde{b}}||_1 \quad s.t \quad ||\boldsymbol{A\tilde{b} - \hat{c}_{CFCOBA}}||_2 \leq \epsilon,
\end{equation}
with $\epsilon$ being an appropriate noise level.
This optimization problem can be solved using various known techniques, such as interior point methods \cite{cands20071} or iterative shrinkage ideas \cite{beck2009fast}, \cite{hale2007fixed}. This optimization problem results in reconstruction of both strong and weak reflectors, as will be shown next through various examples.	 

\section{Evaluation Results}	
\label{sec: EvResults} 
\begin{figure*}[h!]
	\centering
	\includegraphics[width=0.7\textwidth,height=3.2cm]{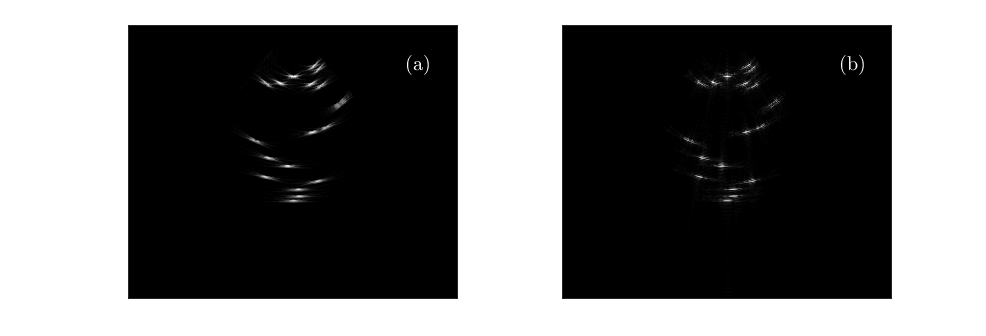}
	\caption{Point scatterers simulated images obtained with (a) DAS (b) Compressed frequency domain COBA - fractal geometry (36-fold reduction).}
	\label{fig:sim}
\end{figure*}
\begin{figure*}[h!]
	\centering
	\includegraphics[width=0.7\textwidth,height=3.2cm]{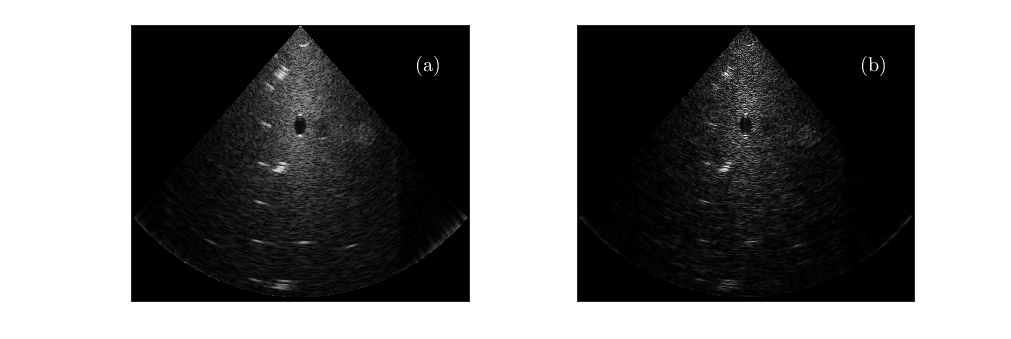}
	\caption{Anechoic cyst phantom images obtained with (a) DAS (b) Compressed frequency domain COBA - fractal geometry (36-fold reduction).}
	\label{fig:phantom}
\end{figure*}
\begin{figure*}[t!]
	\centering
	\includegraphics[width=\textwidth,height=4.6cm]{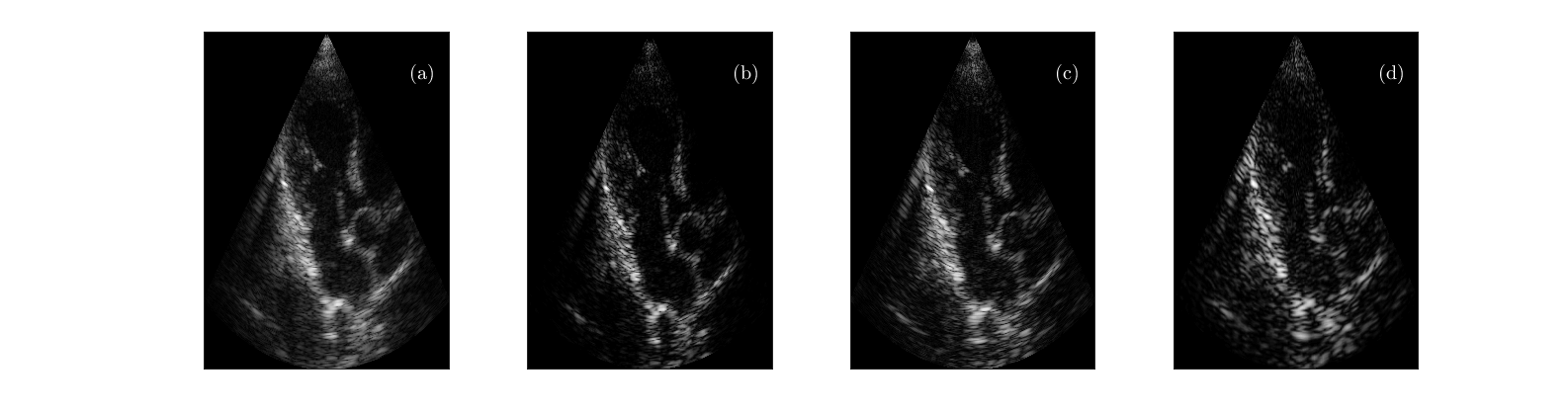}
	\caption{GE US machine cardiac images obtained with (a) DAS (b) COBA - full ULA (c) Non-compressed frequency domain COBA - fractal geometry (35-fold reduction), (d) Compressed frequency domain COBA - fractal geometry (142-fold reduction).}
	\label{fig:GE}
\end{figure*}

\begin{figure*}[t!]
	\centering
	\includegraphics[width=\textwidth,height=4.6cm]{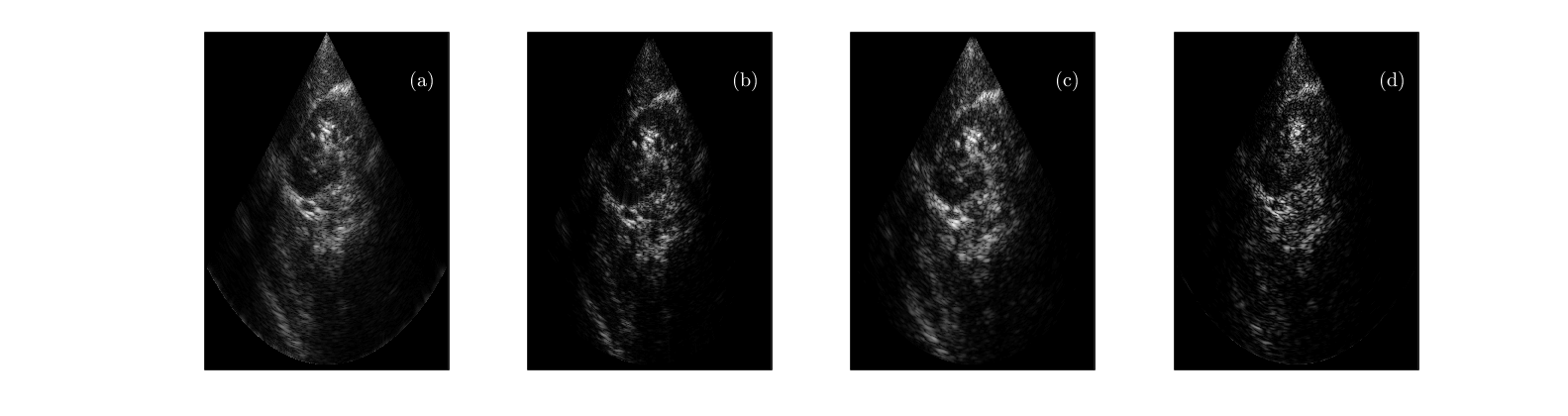}
	\caption{Verasonics US machine kidney images obtained with (a) DAS (b) COBA - full ULA (c) Non-compressed frequency domain COBA - fractal geometry (17-fold reduction), (d) Compressed frequency domain COBA - fractal geometry (36-fold reduction).}
	\label{fig:Ver_kidney}
\end{figure*}

\begin{figure*}[t!]
	\centering
	\includegraphics[width=\textwidth,height=4.6cm]{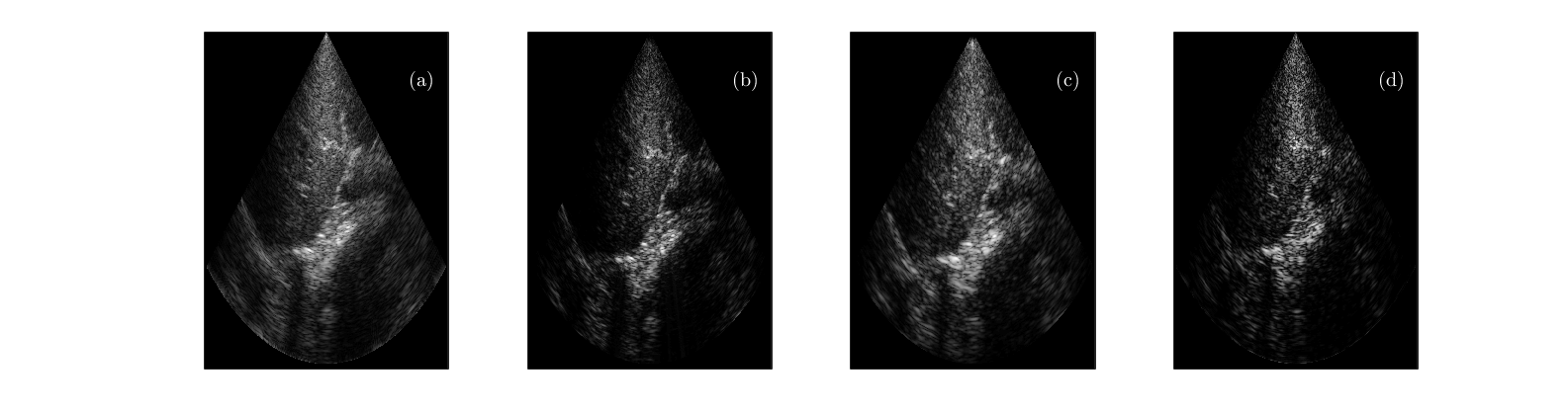}
	\caption{Verasonics US machine liver images obtained with (a) DAS (b) COBA - full ULA (c) Non-compressed frequency domain COBA - fractal geometry (17-fold reduction), (d) Compressed frequency domain COBA - fractal geometry (36-fold reduction).}
	\label{fig:Ver_Liver}
\end{figure*}

\begin{figure*}[t!]
	\centering
	\includegraphics[width=\textwidth,height=4.6cm]{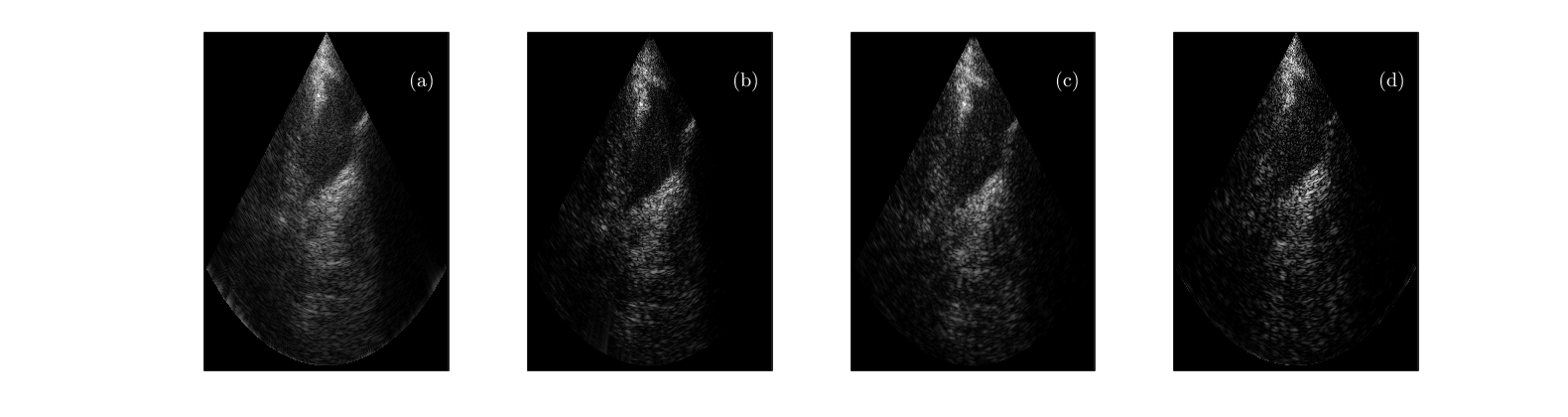}
	\caption{Verasonics US machine bladder images obtained with (a) DAS (b) COBA - full ULA (c) Non-compressed frequency domain COBA - fractal geometry (17-fold reduction), (d) Compressed frequency domain COBA - fractal COBA geometry (36-fold reduction).}
	\label{fig:Ver_bladder}
\end{figure*}
We now demonstrate the performance of the proposed beamforming algorithm in comparison to DAS and the non-compressed version of Fourier domain COBA. The methods are applied to simulated data, tissue mimicking phantoms Gammex 403GSLE and 404GSLE, and RF data acquired from healthy volunteers. For verifying the wide variety of possible usages, we tested the methods on different data sets, each of different body parts. We present here \emph{in vivo} cardiac data, kidney data, liver data, and bladder data.

Acquisition was performed using the GE breadboard ultrasonic scanner and the Verasonics Vantage 256 system. The scans made with the GE US machine were performed using 64 channels phased array probe, with a radiated depth of 16 cm. The probe carrier frequency was 3.4 MHz and the sampling rate was 16 MHz leading to 3328 samples per image line. The scans made with the Verasonics US machine were done using the 64-element phased array transducer P4-2v, with frequency response centered at 2.72 MHz and a sampling rate of 10.8 MHz. For the Verasonics setup, 1920 samples per image line were used. The images were generated using standard steps, including log-compression and interpolation. 

For the fractal array geometry, we used the generator array $\mathbb{G} = \left\lbrace 0,1\right\rbrace $ and array order 4, leading to 15 elements. To perform beamforming in frequency, with the GE imaging setup, we used 400, 100 samples at each of the channels of the fractal array geometry, leading to non compressed and compressed frequency COBA images, respectively. This implies a 142-fold reduction in data size. For the Verasonics machine 480, 230 samples were used to perform non compressed and compressed imaging using the sparse arrays leading to a reduction as high as 36 times less data. To solve \eqref{optimization} we used the NESTA algorithm \cite{nesterov2005smooth}, \cite{becker2011nesta}.

Figs. \ref{fig:sim}-\ref{fig:phantom} show the beamformed US images obtained from point scatterers simulation and phantom scans data sets. In both figures (a) stands for standard DAS and, (b) represents the proposed beamforming method, CFCOBA. The resulting images clearly show that the proposed method outperforms standard methods for US imaging. It can be seen that in the point scatterers simulated data, the points are less blurred and the center of the reflector can be easily noticed. In the phantom setup scans, the strong reflectors are seen as good as in DAS images, and large noise reduction can be observed.

The resulting \emph{in vivo} US images are shown in Figs. \ref{fig:GE}-\ref{fig:Ver_bladder}. The images show four different frames of different body parts, where in each figure (a) corresponds to standard DAS, (b) stands for COBA \cite{cohen2018sparse} with full ULA, (c) stands for non-compressed Fourier domain COBA \cite{mamistvalov2020sparse}, and (d) presents the proposed compressed beamforming method. It can be easily seen that although the images are not identical, the resulting images, using compressed frequency domain COBA, outperform those produced by standard DAS, not only in terms of image resolution and contrast but also in terms of image noise, using two orders of magnitude less data. When comparing the results to the former proposed method, non compressed FDBF, the outcome images do not differ much. These results validate that a significant reduction in both the number of acquiring elements and the sampling rate can be achieved using the proposed technique without degrading image quality and even improving it when compared to traditional DAS.

\section{Conclusion}
\label{sec: conclusion}	
In this paper, we proposed a new beamforming method for high quality B-mode US images. The suggested technique is based on Xampling, frequency domain beamforming, CS, sparse arrays, and convolutional beamforming all implemented efficiently. This combination allows producing a beamformer that can outperform the widely used DAS using much less power and data.

We extended frequency domain COBA presented in \cite{mamistvalov2020sparse} and showed that the convolutional beamforming algorithm can be done directly in the frequency domain. This results in up to 33-fold reduction in sampling rate and 142 times less data combining with sparse convolutional beamforming, without impacting image quality and even improving it. The huge reduction is achieved by sampling the signals at sub-Nyquist rate and using the Xampling mechanism. To reconstruct the signal from partial frequency data, we derived an FRI model for the convolutionally beamformed signal, that resulted in replicas of the square of the known transmitted pulse. This result enabled usage of CS recovery methods.

Finally, we validated our technique both on simulated data and on \emph{in-vivo} channel data of a variety of body parts, acquired by two different US machines, resulting in high quality B-mode US images, using orders of magnitude less data.

Our results prove that the idea of compressed US imaging is feasible for practical use, leading to potential reduction in US cost, power consumption and size, paving the way towards wireless US imaging. 

\section*{Acknowledgment}
The authors would like to thank Dr. Israel Aharony for voluntarily performing the scans which provided the data for testing and evaluating of the proposed method and Sivan Grotas for fruitful discussions.  

\bibliographystyle{ieeetran}
\bibliography{ref}

\end{document}